\documentclass[12pt]{article}
\usepackage{amssymb}
\usepackage{amsmath}
\usepackage{graphicx}
\newcommand{\Mfigure}[1]{
\begin{figure}
\begin{center}
\begin{minipage}{.9\linewidth}
\begin{center}
#1
\end{center}
\end{minipage}
\end{center}
\end{figure}
}
\newlength{\Mwi}   \Mwi   .89\textwidth
\newlength{\Mwii}  \Mwii  8.3cm
\newlength{\Mwiii} \Mwiii 8.3cm
\newlength{\Mwiv}  \Mwiv  8.3cm
\newlength{\Mwv}   \Mwv   8.3cm

\begin{document}
\author{P. Zahn and I. Mertig
\\Institut f\"{u}r Theoretische Physik\\
Technische Universit\"{a}t Dresden, D-01062 Dresden, Germany
}
\title{
c(2x2) Interface Alloys in Co/Cu Multilayers - 
Influence on Interlayer Exchange Coupling and GMR
}
\date{\today}
\maketitle
%
\begin{abstract}
The influence of a c(2x2) ordered interface alloy of 3d transition
metals at the ferromagnet/nonmagnet interface on 
interlayer exchange coupling (IXC), the formation of quantum well
states (QWS) and the phenomenon of Giant MagnetoResistance is
investigated. We obtained a strong dependence of IXC on interface
alloy formation. The GMR ratio is also
strongly influenced. We found that Fe, Ni and Cu alloys at the
interface enhance the GMR ratio for in-plane geometry by nearly a
factor of 2.
\end{abstract}
%
%
\section{Introduction}
In the context of magnetoelectronic applications structural properties
of the interfaces in nanostructured devices have attracted huge
attention. 
Especially, the properties of the ferromagnet/nonmagnet (FM/NM)
interface are expected to be crucial for new phenomena like
oscillatory IXC and GMR that occur in magnetic layered systems. Several
experiments have shown that even in the best homoepitaxially grown
samples interdiffusion at the interfaces exists. The formation of, even
partially ordered surface alloys on a Cu(001) substrate was reported
for, e.g. Mn \cite{flores97}, Fe \cite{johnson94,shen95} and Co
\cite{nouvertne99}.
Recent experiments on Co/Cu spin valve systems have shown the high
sensitivity of the magnetotransport on the interface properties 
by introducing 3d scattering centers at
various positions in the Co and Cu layer ($\delta$-layer doping) 
\cite{marrows00}.
The
Co/Cu(001) system is a model system for magnetism in systems with reduced
dimension \cite{bland94}. The lattice mismatch is rather small
(about 2\%) and the materials are completely immiscible in the bulk
phase. Our investigations are focussed on the electronic properties of
ordered interface alloys in a Co/Cu(001) multilayer system. The
formation of QWS arises from the confinement of the Cu electrons. 
The strong correlation between IXC and quantum well states will be
demonstrated. 
The influence on magnetotransport properties will be discussed.
\section{Method}
The electronic structure of the considered systems was calculated
selfconsistently using a Screened Korringa-Kohn-Rostoker (KKR) method
\cite{szunyogh94,zeller95,zahn99} based on a Green's function multiple
scattering formalism. The screening procedure is an
exact reformulation of the traditional KKR method \cite{korringa47,kohn54} to 
accelerate and simplify the numerical solution of the Dyson equation
for the Green's function. 
The numerical effort scales
linearly with the size of the unit cell for systems with a prolonged supercell.
The systems were modelled as superlattices with a complex unit cell
periodically repeated in 3 
dimensions. That means, we consider infinite multilayer structures
and exclude the influence of surfaces. The Co layer thickness was
fixed to 9 monolayers (ML), $\approx 1.6 nm$ following the experimentally
investigated systems.
We neglect all lattice relaxations at the
interfaces. All atomic positions are fixed to an ideal fcc lattice
with a lattice constant of $6.76~ a.u.$. 
The chosen lattice constant lies between the bulk lattice constant of
magnetic fcc Co ($6.7~a.u.$) and fcc Cu ($6.83~a.u.$).
We use spherical potentials in the atomic sphere approximation
(ASA). The exchange and correlation part is described by means of the
local spin density approximation in the parametrization of Vosko, Wilk and Nusair
\cite{vosko80}. The angular momentum expansion of the Green's function was
treated up to $l_{max}=3$ and for the charge density up to
$l_{max}=6$. To simulate the c(2x2) ordered interface alloy we replace
every second atom in the Co interface layer by a 3d
element (V, Cr, Mn, Fe, Co, Ni, and Cu, resp.), which causes an
increase of the in-plane unit cell by a factor of 2. 
\section{Stability of ordered Interface Alloy}
It was shown by several authors
\cite{bruno95,kudrnovsky97a,levy98,schad99,zahn98} that the structural
properties of the interfaces play an important role for the
interlayer exchange coupling and GMR. The amplitude of IXC and the GMR
ratio can change drastically. We demonstrate that the structural
properties of the interfaces are 
crucial for magnetotransport properties in multilayer systems.
The interface structure depends strongly on growth conditions during
the preparation process. Besides mesoscopic roughness and
interdiffusion the formation of an ordered interface alloy may occur.
The paper is focussed on the role of ordered interface alloys.
First of all we will discuss the stability of the interface alloys in
a Co/Cu multilayer. For this purpose we compare the total energy of
three different interface configurations: 0) the ideal atomically flat
interface with c(1x1) symmetry, 1) one intermixed interface layer with
the same number of Co and Cu atoms equally distributed with c(2x2)
symmetry, and 2) two intermixed interface layers with c(2x2)
symmetry. The considered superlattice of type 0 consist of 9 ML Co and
7 ML Cu. For type 1 and 2 the number of Co layers is reduced by the
intermixed layers to keep the Cu thickness constant. The interface
formation energy is defined to be
\begin{equation}
E^f = E^I - n_{Co} E_{Co} - n_{Cu} E_{Cu}
\end{equation}
where $E^I$ denotes the total energy of the superlattice with one of the
above described interface configurations. 
$E^I$ is compared with the total energies of the bulk systems Co and
Cu, respectively. The bulk calculations have been performed at the
same lattice constant used in the supercell calculation.
$n_{Co}$ and $n_{Cu}$ are the numbers
of $Co$ and $Cu$ atoms in the unit cell, respectively.
$E^f$ and all other energies are  normalized to an interface area of
the c(1x1) structure. 
We
obtain the following values: $E^f_0=190 meV$ for the ideal interfaces,
$E^f_1=240 meV$ for one intermixed atomic layer and $E^f_2=410 meV$
for two intermixed interface layers. The energetical most favourable
configuration is the ideal Co/Cu interface. This result is in
line with the inmiscibility of the materials at low temperatures. The
energy difference of $50 meV$ between the ideal and the one intermixed
layer configuration is of the order of the total energy difference
resolution of our method.
Kinetic reasons during the growth process can stabilize the
configuration with one intermixed layer. The formation energy of the
interface with 2 intermixed layers is remarkable larger. 
These results indicate that the c(2x2) structures found at surfaces with
monolayer and submonolayer coverage of $Co$ on $Cu(001)$
\cite{nouvertne99} should remain at interfaces in multilayers, too. 
The magnetic interaction of the Co
layers (IXC) mediated by the Cu electrons is RKKY-like and for this
system about $1 meV$. The magnetic interaction of adjacent
interfaces due to the finite Co
thickness is expected to be of the same order. Both contributions are
negligible in comparison to
the interface formation energy $E^f$.
\section{Local Density of States}%
We discuss the spin dependent local density of states (LDOS) derived from the
diagonal part of the Green's function and the obtained
magnetic moments of the alloy constituents to characterize the
electronic structure
\begin{equation}
n^\sigma(E,{\bf r}) = - \frac{1}{\pi} {\mathfrak Im} \,G^\sigma(E,{\bf r},{\bf r})
\quad . 
\end{equation}
We denote the alloy
constituents by impurities in the sense of a periodic impurity
arrangement in the interface atomic layer.
\Mfigure{
  \includegraphics[width=\Mwi]{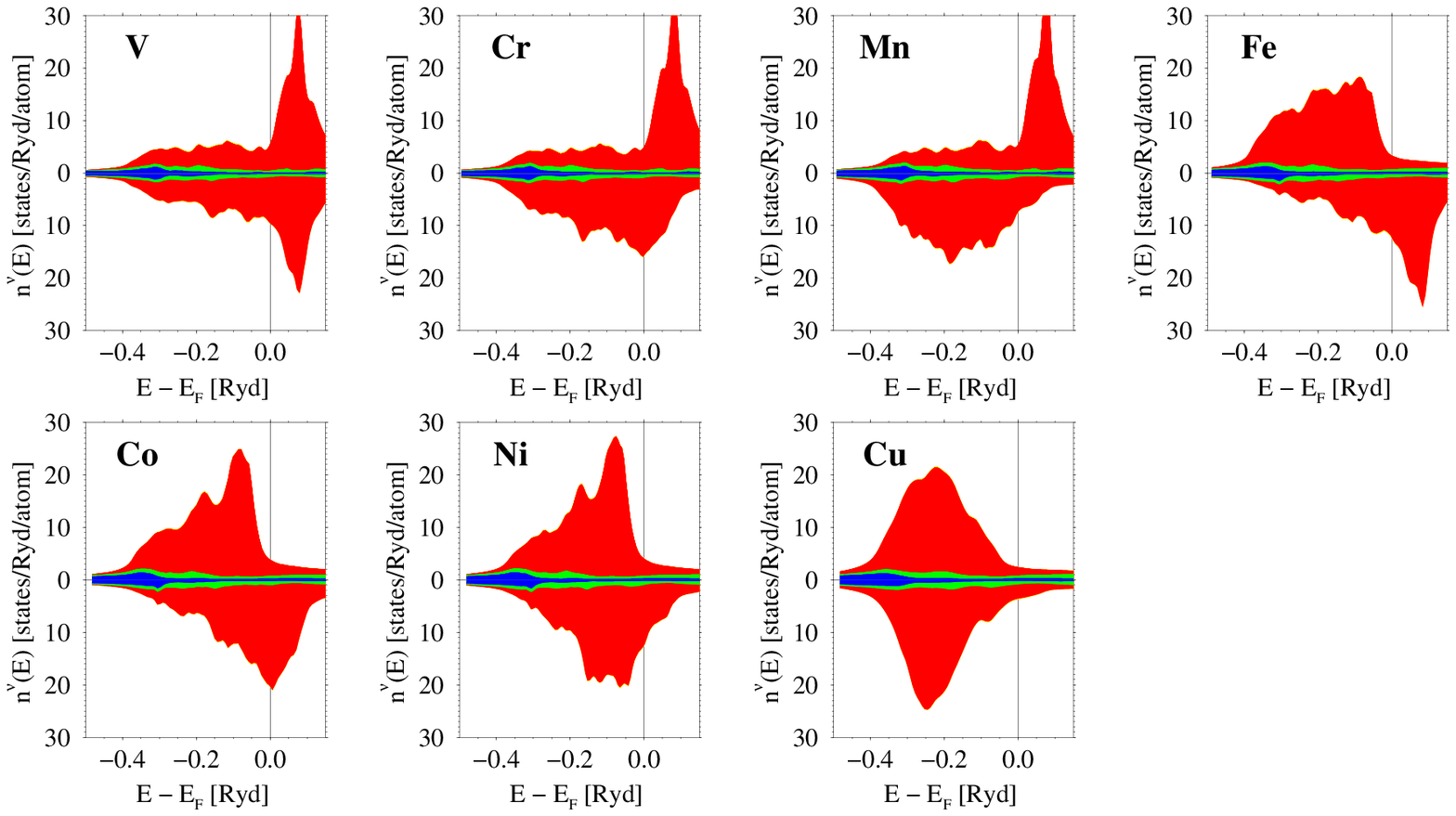}%
  \caption{
Local DOS of the 3d constituents of an ordered c(2x2) interface alloy
Co(3d) at the Co/Cu(001) interface; the blue, green and red areas
correspond to partial s, p and d contributions
  }
  \label{figLDOS}
}
\mbox{Figure \ref{figLDOS}} shows the LDOS in the impurity Wigner-Seitz
sphere for both 
spin directions. The energy is given relative to the Fermi level which
is fixed by the underlying $Co/Cu$ superlattice. 
The system with the ideal interfaces corresponds to the $Co$
interface alloy and has c(1x1) symmetry. 
The local density of states are
comparable to that obtained for 3d impurities in bulk $Co$
\cite{stepanyuk94}. For $V$, $Cr$ and $Mn$ we obtain a sharp virtual
bound state above the Fermi level. The less attractive Coulomb
potential for the early 3d elements cannot be counterbalanced by a
larger exchange-correlation potential. For the later elements $Fe$,
$Ni$, and $Cu$ the exchange-correlation and Coulomb potential act in
opposite directions for the majority electrons. The majority
d-states remain filled and nearly unchanged and the minority states
are shifted to lower energies and cause a reduced magnetic
moment. This behaviour reflects the differences of the impurity
potentials with respect to the Co potential at the ideal $Co/Cu(001)$
interface. The nearly unchanged majority spin potential along with the
drastic changes in the minority channel have a strong impact on the
magnetotransport properties. This will be discussed later in detail.
\par
\Mfigure{
  \includegraphics[width=\Mwii]{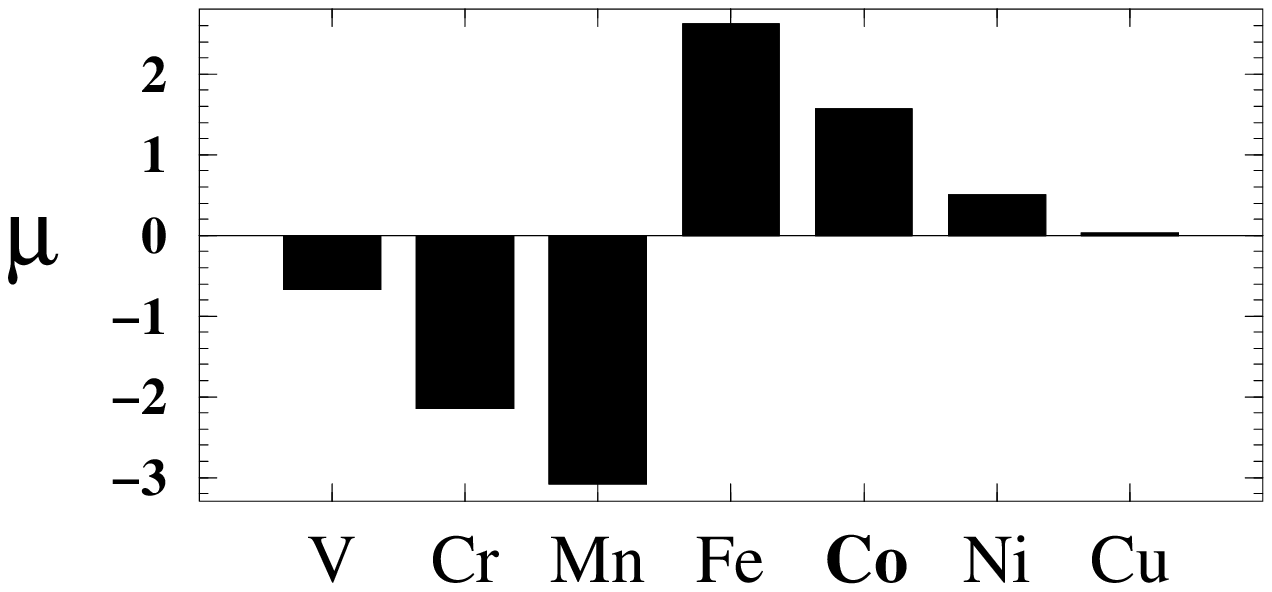}%
  \caption{
Magnetic moments of the 3d constituents of an ordered c(2x2) interface alloy
Co(3d) at the Co/Cu(001) interface
  }
  \label{figMOMENTS}
}
The magnetic moments of the impurities are shown in
\mbox{Fig. \ref{figMOMENTS}}. For the early 3d impurities $V$, $Cr$, and $Mn$ our
calculation yields 
the antiferromagnetic alignment of the impurity moment relative to the
Co moment as ground state. For the second half of the 3d series
$Fe$, $Ni$ and 
$Cu$ the ferromagnetic state is energetically preferred. The size of
the local moment nearly follows Hund's rule with a maximum at half
d-band filling.
\section{Interlayer Exchange Coupling}%
For the calculation of the strength of IXC we apply the frozen potential
approximation (FPA) \cite{oswald85} 
neglecting the double counting parts in the total energy.
We calculate the electronic structure for the
parallel (P) alignment of the Co magnetic moments selfconsistently. The
antiparallel (AP) configuration was obtained by
flipping the magnetization direction in every second Co layer
interchanging the fixed spin dependent ASA potentials. The strength of
the IXC is then given by the single particle contributions to the total
energy
\begin{eqnarray}
\Delta E &=& E^{P}_{SP} - E^{AP}_{SP}\\
E_{SP} &=& - \frac{1}{\pi} \, \sum_\sigma  \,
{\mathfrak Im} \int_{-\infty}^{\mu} dE \int d^3{\bf r} \, (E-\mu) \,
G^{\sigma}(E,{\bf r},{\bf r}) \;-\; \mu N
\end{eqnarray}
with $\mu$ denoting the chemical potential and $N$ the total number of
valence electrons.
The total energy difference $\Delta E $ is proportional to the strength of the
bilinear coupling term which has a cosine dependence on the angle
between adjacent Co layer magnetization directions. 
To determine higher order coupling terms one
has to consider non-collinear magnetic configurations. Usually, the
amplitudes of these contributions are much smaller than the bilinear one
and consequently negligible \cite{blaas99}. 
\par
The strength of IXC varies with the thickness $d$ of the NM
spacer layer. In the asymptotic region at large spacer thicknesses $\Delta E$
is a superposition of oscillatory contributions. The oscillation
periods are determined by stationary points of the spacer Fermi
surface \cite{bruno95,bruno91}. For a $Cu$ spacer in (001)-orientation
2 wellknown contributions exist, a short period from the 
neck region and a long period from the belly region of the 'dogs bone'
orbit. 
\par
In systems with a periodic c(2x2) superstructure the in-plane real space
periodicity is increased. The Fermi surface of Cu has to be folded
down into a smaller Brillouin zone. This is equivalent to the
occurence of Umklapp processes due to different symmetries of spacer
and FM layer. As it was shown by Kudrnovsk\'{y} et al. a new
stationary point on the Cu Fermi surface appears
\cite{kudrnovsky96}. The corresponding period is very short ($\approx
2.15 ML$).
All stationary points and periods obtained in our calculation are
summarized in \mbox{Tab. \ref{tabPERIODS}}.
\begin{table}
\begin{center}
\begin{minipage}{.8\linewidth}
\begin{center}
\caption{Stationary points of $Cu$ in (001) orientation with a c(2x2)
in-plane superstrucure and corresponding oscillation periods of IXC}
\label{tabPERIODS}
\vspace*{3mm}
\begin{tabular}{|l|c|c|}
\hline
& $\boldsymbol{k}^\parallel_i \left[ \frac{\pi}{a} \right]$%
& $\lambda_i \left[ ML \right] $ \\
\hline
$\boldsymbol{k}_1$ & $(0,0)$                  & 5.89 \\
$\boldsymbol{k}_2$ & $(\pm 0.785, \pm 0.785)$ & 2.61 \\
$\boldsymbol{k}_3$ & $(\pm 1,0), (0,\pm 1)$   & 2.15 \\
\hline
\end{tabular}
\end{center}
\end{minipage}
\end{center}
\end{table}
The different oscillatory contributions to the IXC energy are
characterized by their amplitudes $A_i$ and phases $\phi_i$
\begin{equation}
\Delta E = \sum_{i=1}^3 \frac{A_i}{d^2} 
\sin \left( \frac{2\pi}{\lambda_i} d + \phi_i\right)
\quad .
\label{eqCONTRIB}
\end{equation}
These parameters are determined by the coupling of the Cu electrons
to the polarized electrons in the Co layers. They are influenced by
the thickness and chemical composition of the FM layer and the interface
structure. The decay with the square of the spacer thickness arises
from the screening of a 2-dimensional perturbation in a noninteracting
electron gas. 
To determine the oscillatory contributions of IXC we calculated the
coupling strength in dependence on the Cu spacer thickness. The Co
layer thickness was fixed and both interfaces of the Co layers were
decorated with an ordered c(2x2) alloy of the 3d elements. The Cu
thickness was varied in integer steps between 7 and 60 ML. Calculating
the single-particle contribution of the total energy the density of
states is obtained from the diagonal part of the Green's function by a Brillouin
zone integration. To accelerate this integration a Fermi-Dirac
occupation function for the single-particle states with a finite
temperature of $400K$ is used \cite{wildberger95}. This 
smearing of the Fermi edge results in an exponential damping of the
coupling energy for spacer thicknesses much larger than a critical
value L \cite{bruno91}. The decay length L is about 20 ML for
the used temperature, spacer material and crystal orientation. 
\par
\Mfigure{
  \includegraphics[width=\Mwiii]{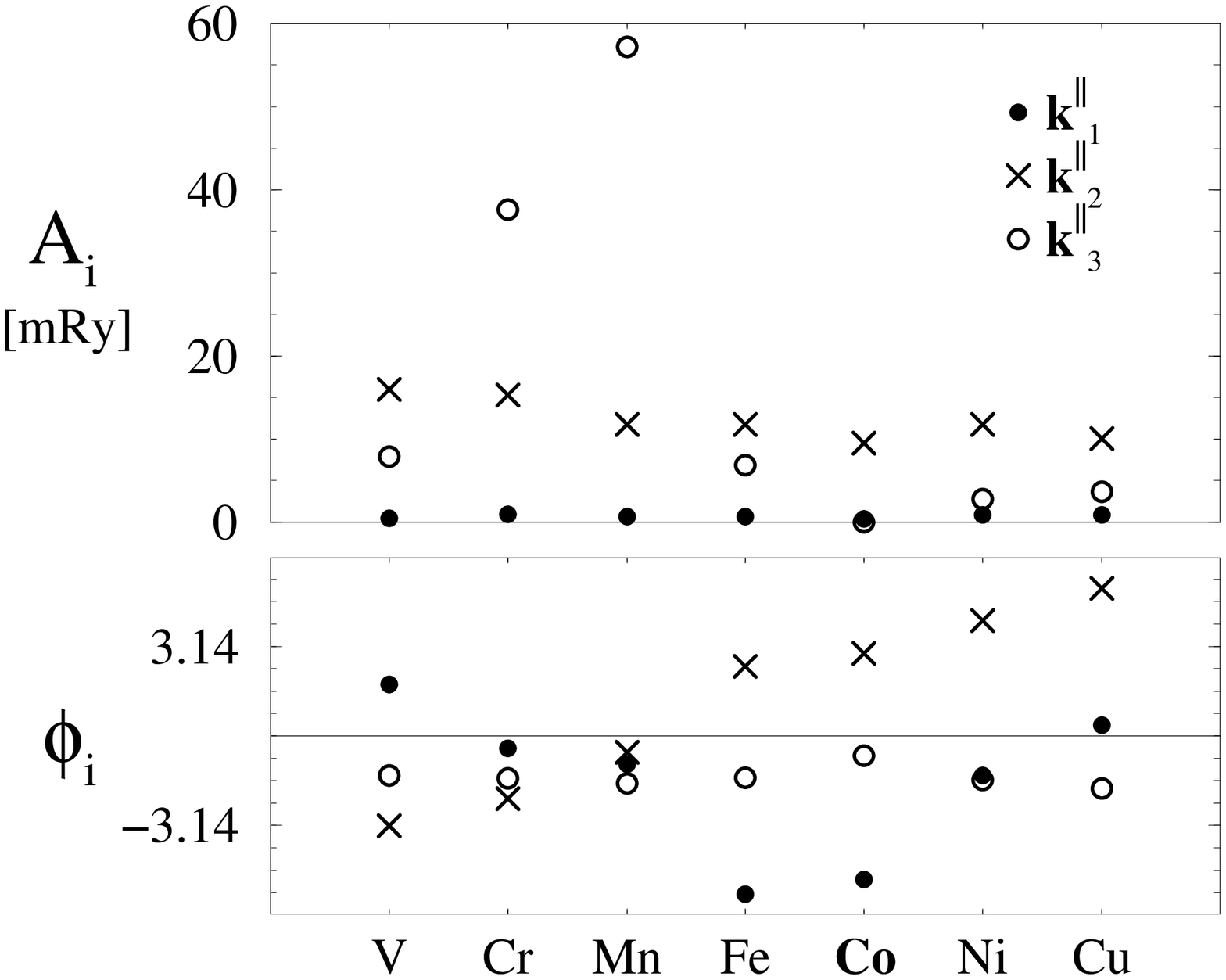}%
  \caption{
Amplitudes $A_i$ and phases $\phi_i$ of the oscillatory contributions to IXC
(\mbox{Eq.~\ref{eqCONTRIB}}) in dependence on constituents for ordered
c(2x2) alloys at the Co/Cu interface
  }
  \label{figAMPLITUDES}
}
From a discrete Fourier transformation at the known periods
$\lambda_i$ we obtain the amplitudes and phases of the different
contributions. \mbox{Figure \ref{figAMPLITUDES}} shows the amplitudes $A_i$ and
phases $\phi_i$ for the three contributions. The amplitude
$A_1$ of the long period oscillation is very small for all considered
systems including the
ideal Co/Cu superlattice. The amplitude $A_2$ of the short period
oscillation dominates the coupling energy for nearly all systems
and varies by about 30\%.
The largest variations were found for the 
amplitude $A_3$ of the oscillation caused by the superstructure. 
Due to symmetry arguments the amplitude $A_3$ vanishes for
the system with ideal interfaces. This sytem has a c(1x1) in-plane
symmetry and $\boldsymbol{k}_3$ is not a stationary point at the Cu Fermi
surface. 
The amplitude $A_3$ dominates for the systems with $Cr$ and $Mn$
interface alloys and is about 3 times larger than the short period
amplitude $A_2$.
To elucidate the origin of this large contribution to the interlayer
coupling strength
we focus on the formation of quantum well states in
the multilayer structure.
\section{Quantum Well States}%
The description of the IXC in the RKKY picture was succesfully applied
to explain the oscillatory contributions and the periods
\cite{bruno91}. To
determine the phases and amplitudes of the oscillations the resulting
electronic structure of the multilayer has to be examined. Crucial
for the IXC is the occurence of spin polarized localized states (QWS) in the
spacer layer near the Fermi level. The boundary conditions caused by
the spin dependent potential give rise to the formation of QWS. Changing the
relative orientation of the magnetization direction of adjacent Co
layers the energy of the
QWS is shifted. In case quantum well states are shifted across the Fermi level the
redistribution of electrons causes a change in the total energy and
leads to the preference of one magnetic configuration to the other.
\par
To elucidate the drastic changes in the amplitudes of the IXC
contributions in dependence on the alloy constituent we investigated
the $\boldsymbol{k}$-projected partial LDOS (kpLDOS) at the Fermi level. We
obtain this quantity by integrating the diagonal part of the
one-electron Green's function in real space over the Wigner Seitz sphere and in
reciprocal space over $k_\perp$ perpendicular to the plane
\begin{equation}
n_l^\nu(\boldsymbol{k}_\parallel) = - \frac{1}{\pi} \, \sum_m 
\int d{k}_\perp \; {\mathfrak Im} \,
G^{\nu\nu}_{LL}(z,\boldsymbol{k}_\parallel,{k}_\perp)
\quad .
\end{equation}
$L=(l,m)$ is a short hand notation for orbital and magnetic quantum
number, $\nu$ is the site index in the unit cell and $z$ the Fermi
energy $\epsilon_F$ with a small complex part $\Gamma$ of about
$0.2eV$ to achieve 
a better convergence in the $k_\perp$ integration.
The kpLDOS is related to the probability amplitude of the
eigenstates at the Fermi level
\begin{align}
n_l^\nu(\boldsymbol{k}_\parallel) &=  \frac{1}{\pi} \sum_m 
\int dE \int dk_\perp \; 
\left| \Psi^{\nu L}_{(\boldsymbol{k}_\parallel,{k}_\perp)} (E)
\right|^2
\frac{\Gamma}{(E - \epsilon_F)^2 + \Gamma^2} \\
&\approx \sum_{{k}_perp} \delta(E - \epsilon_F)
\left| \Psi^{\nu L}_{(\boldsymbol{k}_\parallel,k_\perp)} (\epsilon)
\right|^2
\quad .
\end{align}
The typical probability amplitude for confined states in a layered
strucure can be obtained for a free electron model with potential wells
in growth direction. From the boundary condition that the
wave function decreases exponentially in the potential barrier one
obtains an oscillatory probability amplitude inside the potential
well. The oscillation period is given by the wave vector of the free
states without any barrier. In the NM/FM multilayer system the
potential barrier corresponds to the FM layers and the states in the
NM spacer are confined in the resulting potential. The oscillation of
the probability amplitude is given by the unperturbed spacer states. A
strong oscillatory variation of the probability amplitude inside the
the spacer indicates a strong confinement and a large
contribution to the IXC from the considered stationary
$\boldsymbol{k}_\parallel$ vector. 
\Mfigure{
  \includegraphics[width=\Mwiv]{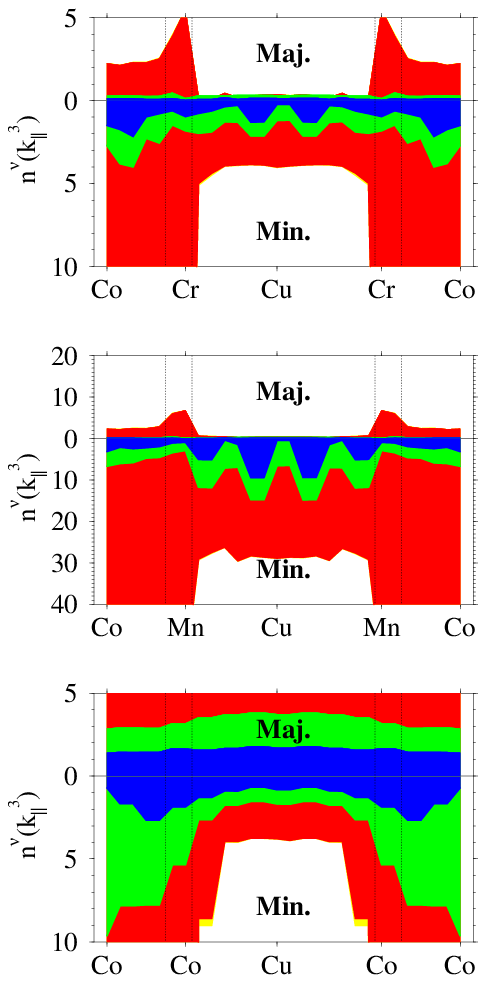}%
  \caption{
Spin dependent $\boldsymbol{k}$-projected Local DOS at the stationary point
$\boldsymbol{k}_\parallel^3$ for the $Cr$ alloyed (top), the $Mn$ alloyed
systems (center) and the system with ideal interfaces (bottom); the
blue, green and red areas 
correspond to partial s, p and d contributions 
  }
  \label{figKPLDOS}
}
\mbox{Figure \ref{figKPLDOS}} shows the
$\boldsymbol{k}$-projected partial LDOS at the stationary
$\boldsymbol{k}_\parallel^3$ point for the system with ideal interfaces, the
$Cr$, and the $Mn$ alloy. 
The spin-dependent kpLDOS is shown for every atomic layer for the
P alignement of the $Co$ magnetic moments. We remind the reader that every
atomic layer is represented by 2 atoms. On the left and right are the
Co layers, the dashed vertical lines mark the alloyed Co interface
layers and in between is the Cu spacer.
For the ideal system the states at
$\boldsymbol{k}_\parallel^3$ have a nearly constant amplitude in the Cu
spacer and a strong hybridization with the Co states for both spin
directions. In contrast, the minority states in the $Cr$ and $Mn$
alloyed sytems show strong indications of quantum confinement in the
spacer region. The s- and d-angular momentum contributions show clear
oscillations with a period of about $2 ML$.
This coincides with the Fermi wave vector of $Cu$ at
$\boldsymbol{k}_\parallel^3$ which determines the oscillation period of the
superstructure contribution to IXC. At the position of the
stationary point $\boldsymbol{k}_\parallel^2$ we found for all the considered
systems confined states in the $Cu$ spacer for minority and majority
spin electrons.
\section{Giant MagnetoResistance}%
For the calculation of transport properties we use Mott's two-current
model \cite{mott64} neglecting spin flip scattering. For the
considered systems assuming non-Kondo-impurities only this assumption
should be reasonable. The linearized Boltzmann equation in relaxation
time approximation was solved. One yields the spin dependent
conductivity $\sigma^\sigma$ from a Fermi
surface integral weighted with an isotropic relaxation time neglecting
the anisotropy of the scattering operator
\begin{equation}
\sigma^\sigma = \frac{e^2}{V} \; \tau \; 
\sum_{\boldsymbol{k}} \delta ( \epsilon_{\boldsymbol{k}}^\sigma - \epsilon_F) 
\boldsymbol{v}_{\boldsymbol{k}}^\sigma \otimes \boldsymbol{v}_{\boldsymbol{k}}^\sigma
\qquad .
\end{equation}
The superscript $\sigma$ denotes the spin direction and $V$ the
crystal volume. A modified tetrahedron method was used to obtain the
Fermi surface integrals \cite{lehmann72,zahn98a}. Assuming the same
spin independent relaxation times $\tau$ for both magnetic configurations of
the Co moments (P,AP) this quantity does not enter the expression for
the GMR ratio. The GMR ratio is defined in the usual way by
normalizing the resistivity drop in a sufficient large magnetic field
by the saturation resistivity. In terms of conductivities for the P
and AP configurations this is
\begin{equation}
R = \frac{\sigma^P}{\sigma^{AP}} \, - \, 1
\qquad .
\end{equation}
\Mfigure{
  \includegraphics[width=\Mwv]{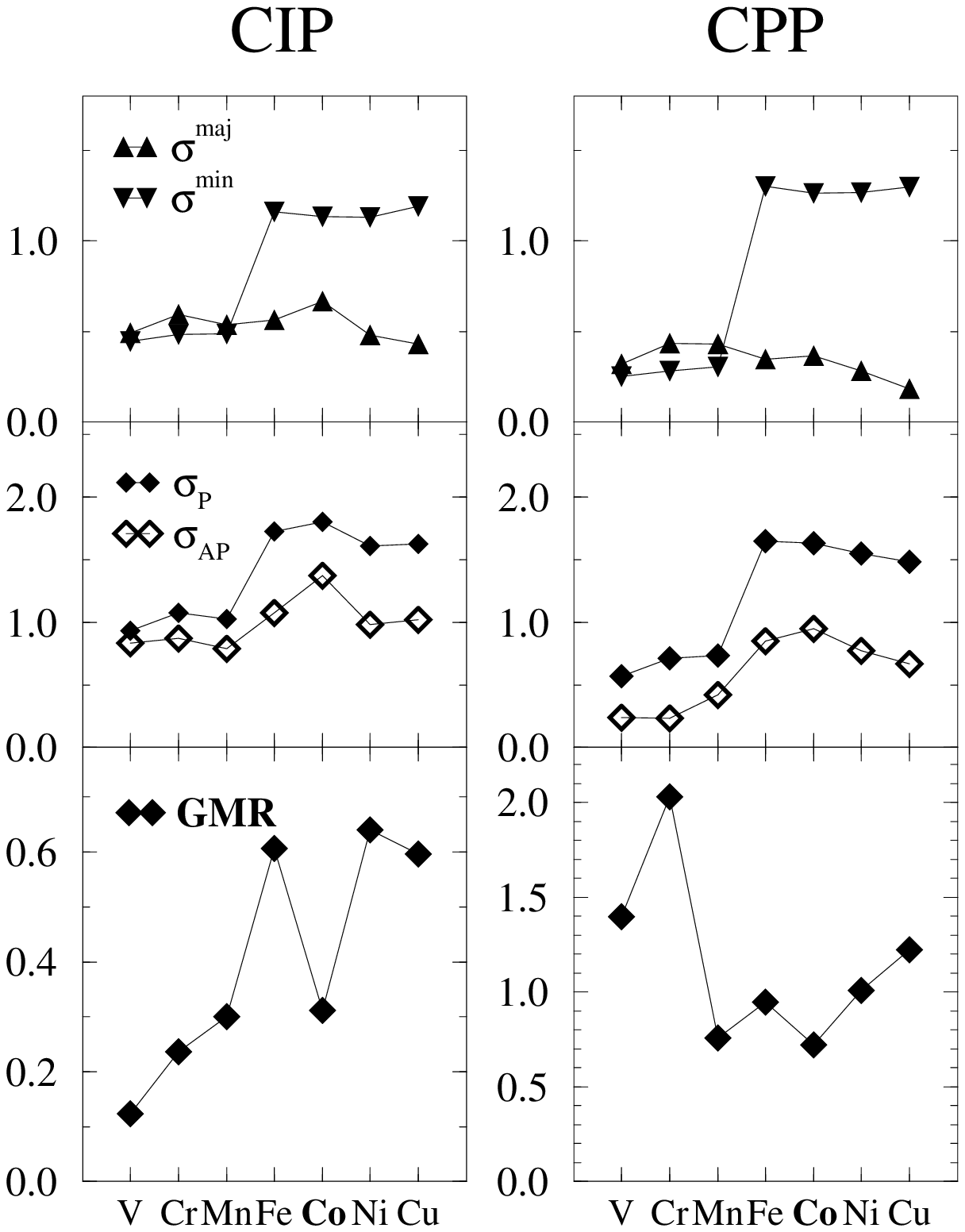}%
  \caption{
Spin resolved conductivities for the P configuration (top panels),
conductivities for P and AP configuration (central panels) and GMR ratio (lower
panels) for CIP and CPP transport in dependence on interface alloy constituent
  }
  \label{figGMR}
}
In \mbox{Fig. \ref{figGMR}} (upper panels) the spin resolved conductivity contributions for the
P configuration are shown in dependence on the interface alloy
constituent. The conductivities are calculated for current
in-plane (CIP) and perpendicular-to-plane (CPP) geometry. For defects
with a moment parallel to the Co magnetization a large spin
anisotropy occurs. The conductivity is dominated by the fast majority
channel, especially pronounced in the CPP geometry. The occurence of one
dominating spin channel in the P configuration is essential to
establish a large GMR ratio. In $Co/Cu$ systems the majority potential
for both materials is nearly equal whereas the minority potential has
strong quantum well modulations. For the AP configuration the
effective potential for both spin channels is a mixing of majority and
minority potentials and causes averaged conductivities. The
same effect is produced by defects with an antiparallel moment relative
to the $Co$ moment causing a strong perturbation of the majority
potential even in the P configuration. 
\par
In \mbox{Fig. \ref{figGMR}} (central panel) the total conductivities for the P and AP
confuguration are compared. The relative difference determines the GMR
ratio. We obtain a CIP-GMR ratio
of $31 \%$ for the system with ideal interfaces.
For the alloys of $Fe$, $Ni$, and $Cu$ with a moment parallel to the
$Co$ moment the value is increased by nearly a factor of 2. In
the contrary, for systems with alloys of $V$, $Cr$, and $Mn$ with antiparallel
moment the CIP-GMR ratio is reduced systematically. As a result, we
can conclude the following
empirical rule for the influence of interface alloys on CIP
transport. Magnetic ordered structures with moments parallel to the
ferromagnetic layers increase the GMR ratio whereas alloy constituents
with antiparallel moments decrease the ratio.

In CPP geometry the GMR ratio changes by about 60\% for alloys with
parallel moments. In the case of antiparallel moment of the alloy
constituent the spin asymmetry of the conductivity in the P
configuration changes the sign and the conductivity is dominated by
the minority channel. Additionally, the conductivity for AP
configuration decreases rapidly for the earlier 3d elements and causes
a strong enhancement of GMR for $V$ and $Cr$ interface alloys.
\section{Summary}%
We demonstrated that the electronic structure of a Co/Cu(001)
multilayer systems is strongly influenced by the formation of an
ordered interface 
alloy. The correspondence between the formation of quantum well states
and the 
strength of oscillatory IXC in dependence on the interface structure
was demonstrated.
We found a third oscillatory contribution to IXC arising from the
c(2x2) interface alloy symmetry. The amplitude of this contribution
varies strongly in dependence on the alloy constituent. Due to the
very short oscillation period the experimental confirmation of this
contribution would be a strong challenge.
The GMR ratio is mainly determined by the coherent potential landscape
of the multilayer. The resistivity drop is caused by the
occurence of a fast and slow channel in the P configuration. The GMR
ratio decreases rapidly with the vanishing spin asymmetry of the 
conductivity in P configuration due to alloy constituents with an
antiparallel moment. 
\par
We would like to thank G. G{\"u}ntherodt, P.H. Dederichs, and
J. Binder for stimulating discussions. P.Z. ackknowledges the support
by the German BMBF-Leitprojekt "Magnetoelektronik" (contract
13N7379). 
%

%
%

\begin{thebibliography}{99}
\bibitem[1]{flores97}
T. Flores, S. Junghans, and M. Wuttig,
Surf. Sci. {\bf 371}, 1 (1997).
%
\bibitem[2]{johnson94}
K.E. Johnson, D.D. Chambliss, R.J. Wilson, and S. Chiang,
Surf. Sci. {bf 313}, L811 (1994).
%
\bibitem[3]{shen95}
J. Shen, J. Giergel, A.K. Schmid, and J. Kirschner,
Surf. Sci. {\bf 328}, 32 (1995).
%
\bibitem[4]{nouvertne99}
F. Nouvertne, U. May, A. Rampe, M. Gruyters, U. Korte, R. Berndt, and
G. G{\"u}ntherodt,
Surf. Sci. {\bf 436}, L653 (1999).
%
\bibitem[5]{marrows00}
C.H. Marrows and B.J. Hickey, cond-mat/0005073
%
\bibitem[6]{bland94}
J.A.C. Bland and B. Heinrich (eds.),
{\it Ultrathin Magnetic Structures I+II}, Springer, Berlin, (1994).
%
\bibitem[7]{szunyogh94}L. Szunyogh, B. \'{U}jfalussy, P. Weinberger,
and J. Koll\'{a}r, Phys. Rev. B {\bf 49}, 2721 (1994).
%
\bibitem[8]{zeller95}
R. Zeller, P.H. Dederichs, B. \'{U}jfalussy, L. Szunyogh, and
P. Weinberger, Phys. Rev. B {\bf 52}, 8807 (1995).
%
\bibitem[9]{zahn99}
P. Zahn, I. Mertig, R. Zeller, and P.H. Dederichs,
Phil. Mag. B {\bf 78}, 411 (1998).
%
\bibitem[10]{korringa47}J. Korringa, Physica {\bf 13}, 392 (1947).
%
\bibitem[11]{kohn54}W. Kohn and N. Rostoker, Phys. Rev. {\bf 94}, 1111 (1954).
%
\bibitem[12]{vosko80}S. H. Vosko, L. Wilk, and M. Nusair,
Can. J. Phys. {\bf 58}, 1200 (1980).
%
\bibitem[13]{bruno95}
P. Bruno, Phys. Rev. B {\bf 52}, 411 (1995).
%
\bibitem[14]{kudrnovsky97a}
J. Kudrnovsk\'{y}, V. Drchal, R. Coehoorn,
M. \v{S}ob, and P. Weinberger, Phys. Rev. Lett. {\bf 78}, 358 (1997).
%
\bibitem[15]{levy98}
P.M. Levy, S. Maekawa, and P. Bruno,
Phys. Rev. B {\bf 58}, 5588 (1998).
%
\bibitem[16]{schad99}
R. Schad, P. Belien, G. Verbanck, V. V. Moshchalkov, Y. Bruynseraede,
H. E. Fischer, S. Lefebvre, and M. Bessiere,
Phys. Rev. B {\bf 59}, 1242 (1999).
%
\bibitem[17]{zahn98}
P. Zahn, J. Binder, I. Mertig, R. Zeller, and P.H. Dederichs,
Phys. Rev. Lett. {\bf 80}, 4309 (1998).
%
\bibitem[18]{stepanyuk94}
V.S. Stepanyuk, R. Zeller, P.H. Dederichs, and I. Mertig,
Phys. Rev. B {\bf 49}, 5157 (1994).
%
\bibitem[19]{oswald85}A. Oswald, R. Zeller, J. Braspenning, and
P.H. Dederichs, J. Phys. F {\bf 15}, 193 (1985).
%
\bibitem[20]{blaas99}
C. Blaas, P. Weinberger, L. Szunyogh, J. Kudrnovsky, V. Drchal,
P.M. Levy, and C. Sommers,
Eur. Phys. J. B {\bf 9}, 245 (1999).
%
\bibitem[21]{bruno91}
P. Bruno and C. Chappert,
Phys. Rev. Lett. {\bf 67}, 1602, 2592 (E) (1991)
%
\bibitem[22]{kudrnovsky96}
J. Kudrnovsk\'{y}, V. Drchal, C. Blaas, I. Turek, and P. Weinberger,
Phys. Rev. Lett. {\bf 76}, 3834 (1996).
%
\bibitem[23]{wildberger95}
K. Wildberger, P. Lang, R. Zeller, and P.H. Dederichs,
Phys. Rev. B {\bf 52}, 11502 (1995).
%
\bibitem[24]{mott64}N.C. Mott, Adv. Phys. {\bf13}, 325 (1964)
%
\bibitem[25]{lehmann72}G. Lehmann and M. Taut,
Phys. Stat. solid (b) {\bf 54}, 469 (1972).
%
\bibitem[26]{zahn98a}
P. Zahn, PhD Thesis, TU Dresden (1998),
\\ http://www.phy.tu-dresden.de/physik/publik/habildiss.htm .
%
\end{thebibliography}
\end{document}